\documentclass[aps,pre,preprint,showpacs,eqsecnum]{revtex4-1}
\usepackage[centertags]{amsmath}
\usepackage{amssymb,amsthm,graphicx,epsfig,subfigure,mathrsfs,wasysym}

\begin{document}
\title{Fluctuation corrections on thermodynamic functions; finite size effect}
\author{Sudarson Sekhar Sinha, Arnab Ghosh and Deb Shankar Ray{\footnote {e-mail address: pcdsr@iacs.res.in
}}}
\affiliation{Indian Association for the Cultivation of Science, Jadavpur, Kolkata 700032, India.\\
}
\vspace{1.5cm}
\begin{abstract}
\vspace{0.3cm}
The explicit thermodynamic functions, in particular, the specific heat of a spin system interacting with a spin bath which exerts finite dissipation on the system are determined.
We show that the specific heat is a sum of the products of a thermal equilibration factor that carries the temperature dependence and a dynamical correction factor, characteristic of the dissipative energy flow under steady state from the system.
The variation of specific heat with temperature is accompanied by an abrupt transition that depends on these dynamical factors characteristic of the finite system size.

\end{abstract}
\maketitle
\newpage

\section{Introduction} \label{s1}
The development of quantum theory of Brownian motion laid the foundation of non-equilibrium statistical mechanics of a system coupled to its environment\cite{hanggi_rmp,grabert}.
This has significantly affected the course of research in quantum optics, condensed matter physics and chemical dynamics, over the last several decades\cite{louisell,nitzan}.
Although its major thrust lies on calculation of non-equilibrium properties in terms of relevant correlation functions, a few interesting attempts have been made in evaluating thermodynamic quantities of small quantum system coupled to its environment which exerts a finite dissipation on the system.
For example, the free energy of a quantum oscillator interacting via dipole coupling to black-body radiation field modes has been calculated\cite{ford_prl} to obtain a temperature-dependent shift in the free energy resulting in modification of Planck's formula.
The effects of dissipation on thermodynamic functions have been explored for quantum oscillators in contact with Ohmic and non-Ohmic reservoirs\cite{hanggi_njp} and for the magnetic moment of an electron gas\cite{ford_prb}.
It has also been shown that while a classical free particle does not obey third law of thermodynamics\cite{hanggi_app} its coupling to a thermal reservoir renders quantum nature and allows recovery of third law\cite{ford_prl,hanggi_njp,ford_prb,hanggi_app}.
That quantum dissipation helps to ensure the validity of third law has been a major finding for a dissipative quantum oscillator\cite{hanggi_njp}, for which the specific heat at low temperature exhibits power law behaviour.

The focus of the present paper is the calculation of thermodynamic functions, particularly the specific heat of a spin-$\frac{1}{2}$ particle coupled to a spin bath.
To bring forth the discussion in an appropriate perspective we first note the following points.
When a small system is made open by coupling it to an environment the system experiences a dissipative flow of energy into the environment.
It is easy to anticipate that such a flux would give rise to a dynamical correction over the usual thermodynamic   contribution responsible for the temperature dependence of the system, a situation reminiscent of Kramer's theory of activated rate processes\cite{hanggi_rmp,grabert}.
In the later case the rate constant is essentially a product of the equilibrium contribution due to transition state theory and a dynamical factor due to the frictional coefficient characteristic of the thermal bath\cite{nitzan}.
It is possible to realize the nature of dynamical corrections over the thermodynamic functions by spatially restricting the system size when the fluctuations of the internal energy of the system can be quite significant.
In this paper we focus on the dynamics of an open spin system experiencing spin-bath induced dissipative effects\cite{ferrer} as an underlying microscopic description of this behaviour.
To be specific, the behaviour of specific heat for small systems can be analysed by examining the finite size effects of the system\cite{tang, kodama_prb, viswanatha, stichtenoth}.
The strategy for calculation is based on the free energy of the spin system coupled to the spin bath\cite{sinha_pre,ghosh_jcp} minus the free energy of the spin bath in absence of the system.
The desired free energy of the interacting spin system turns out to be an integral over the free energy of a single system multiplied by a density of states related to the susceptibility\cite{ford_jsp} derived explicitly from the associated c-number quantum Langevin equation.
For implementing the scheme we have employed the spin coherent state representation\cite{radcliffe} of our proposed Hamiltonian of the spin-spin-bath model to cast the c-number Hamiltonian into an oscillator-oscillator bath model.
The key point is the use of Holstein-Primakoff transformation\cite{holstein} which sets up a mapping between spin and boson operators.
Interestingly this allows us to recover the traditional form of quantum Langevin equation (its noise correlation, however, is guided by the spin characteristics of the bath) and consequently the associated susceptibility.
We derive the explicit analytic expression of the specific heat and its high and low temperature limits, which carry the signature of dissipative effect of the environment.
It has been shown that specific heat can be expressed as a sum of the products of an equilibration factor that contains the usual temperature dependence and a dynamical correction factor characteristic of the dissipative flow of energy from the system.
Finally, our analysis is correlated with finite size effects on the variation of specific heat with temperature characterized by abrupt transition as one passes from low to high temperature regime.

The outline of the paper is as follows:
We introduce the spin-spin-bath model and the associated Hamiltonian followed by its spin coherent state representation in Sec II.
The basis of this analysis is the Holestein-Primakopff transformation which further allows us to map the problem to an oscillator-oscillator-bath Hamiltonian.
The c-number quantum Langevin equation is a direct consequence of this mapping.
In Sec III we calculate the thermodynamic functions, free energy and specific heat for the interacting spin system and the high and low temperature limits of the derived expressions and correlated them with dynamical factors characterized by system size.
The paper is concluded in Sec IV.

\section{Quantum dynamics in a spin bath} \label{s2}
\subsection{The Model}
To set up the problem of quantum dissipation of a two-level system in a sea of a two-level atoms, we consider the following Hamiltonian:

\begin{equation} \label{e2:1}
\hat{H}=\hslash\omega_0\hat{S}_z+\hslash\sum_k\omega{_k}\hat{S}_{zk} +\hslash\sum_kg{_k}\left\lbrace \frac{g'_k}{\omega_k}\hat{I}-\left(\hat{S}_k^\dagger+\hat{S}_k\right)\left(\hat{S}^\dagger+\hat{S}\right)\right\rbrace
\end{equation}
Here the first term is the Hamiltonian for the system, which is specified by the Pauli population difference operator $\hat{S}_z$, the other operators being $\hat{S}^\dagger$, $\hat{S}$ and $\hat{I}$, where $\hat{S}^\dagger$ and $\hat{S}$ are the usual raising and lowering operators and $\hat{I}$ is the identity operator.
The second term corresponds to the reservoir Hamiltonian; the bath operators are denoted by $\hat{S}_k^\dagger$, $\hat{S}_k$ and $\hat{S}_{zk}$ (the subscript refers to the k-th atom of the bath).
The last term represents the interaction between the system and the bath atoms.
$\omega_0$ and $\omega_k$ are the characteristic frequencies of the two-level system and the k-th two-level bath atom.
$g_k$ and $g'_k$ are the coupling constant and a frequency scale factor (this is a necessity for correct normalization of the spin coherent states used in the next section).
The Pauli operators for the system follow the usual commutation relations as given by:
\begin{equation} \label{e2:2}
[\hat{S}^\dagger,\hat{S}]=2\hat{S}_z; [\hat{S}^\dagger,\hat{S}_z]=-\hat{S}^\dagger; [\hat{S},\hat{S}_z]=\hat{S}
\end{equation}

We have $\hat{S}^\dagger = \hat{S}_x + i\hat{S}_y$, and $\hat{S} = \hat{S}_x - i\hat{S}_y$.
The commutation and anti-commutation rules between the spin-$ \frac{1}{2} $ operators are given by,
\begin{eqnarray} \label{e2:3}
\hat{S}^\dagger\hat{S} - \hat{S}\hat{S}^\dagger &=& 2\hat{S}_z \nonumber\\
\hat{S}^\dagger\hat{S} + \hat{S}\hat{S}^\dagger &=& 1
\end{eqnarray}
So, $\hat{S}_z = \hat{N}-\frac{1}{2}$, where $\hat{N}$ is the number operator $\hat{S}^\dagger\hat{S}$.
The anti-commutation relation in Eq. \eqref{e2:3} has the immediate consequence that spin-$ \frac{1}{2} $ particle or two-level atom obey Fermi-Dirac statistics.
In view of Eqs. \eqref{e2:1} and \eqref{e2:3} the system Hamiltonian $\hat{H}_s$ may also be expressed as $\hat{H}_s=\hslash\omega_0\hat{S}_z = \hslash\omega_0(\hat{N}-\frac{1}{2})$, while the eigen-value equation for the number operator may be written as $\hat{N}|n\rangle = n|n\rangle$ with n=0,1.
The general state with no quanta is denoted by $|0\rangle$ which satisfies $\hat{N}|0\rangle=0$ and $\hat{H}_s|0\rangle = \hslash\omega_0(\hat{N}-\frac{1}{2})|0\rangle=-\frac{\hslash\omega_0}{2}|0\rangle$ and the state with one quantum is denoted by $|1\rangle$ which obeys $\hat{N}|1\rangle=1$ and $\hat{H}_s|1\rangle=\hslash\omega_0(\hat{N}-\frac{1}{2})|1\rangle=\frac{\hslash\omega_0}{2}|1\rangle$.
Similar relations hold good for bath operators.
Note that unlike harmonic oscillator, the Hamiltonian is asymmetric, both for the system and the bath and the Hilbert space is two-dimensional.

Our object in the next section is to construct a c-number Hamiltonian from Eq.\eqref{e2:1} using spin coherent state representation as introduced by Radcliffe\cite{radcliffe} a couple of decades ago.
Application of spin coherent states are well known in the context of ferromagnetic spin wave, phase transition in Dicke model of super-radiance, equilibrium statistical mechanics of radiation-matter interaction and so on.
For details we refer to Klauder and Skagerstam\cite{klauder}.

\subsection{Spin-spin-bath Hamiltonian in coherent state representation}\label{s4}

Now we return to Eq. \eqref{e2:1} and carry out quantum mechanical averaging with product separable coherent states of the system and the bath at t=0, $|\xi\rangle|\mu_1\rangle|\mu_2\rangle...|\mu_N\rangle$, where $|\xi\rangle$ denotes the coherent state of the two-level system and $|\mu_k\rangle$ corresponds to the coherent state of the k-th bath atom.
The normalized spin coherent state $\vert\mu\rangle$ is defined as,
\begin{equation} \label{e2:4}
\vert\mu\rangle = \left(1+\vert\mu\vert\right)exp(\mu\hat{S})\vert 0\rangle
\end{equation}
where the ground state $\vert 0\rangle$ corresponds to the state with minimal projection $m_s=-S$ and $\mu$ is the c-number.
In using the coherent state representation of the Hamiltonian, we express the identity operator in Eq. \eqref{e2:1} in terms of the system operators;
\begin{equation} \label{e2:5}
\hat{I} = \hat{S}^2 +\hat{S}^{\dagger 2}+\hat{S}\hat{S}^\dagger+\hat{S}^\dagger\hat{S}
\end{equation}
Here we have used the anti-commutation relation of the spin-$ \frac{1}{2} $ or two-level system operators.
The consequence of expressing the identity operator as \eqref{e2:5} will be clear immediately.
The set of c-numbers for the bath degrees of freedom $\left\lbrace \mu_k(0),\mu_k^*(0) \right\rbrace $, k = 1,2,...$\infty$, obtained from the different matrix elements, as discussed in \cite{ghosh_pre}, can then be expressed in terms of momenta $\left\lbrace\beta_k\right\rbrace $ and coordinates $\left\lbrace\alpha_k\right\rbrace $ of  bath oscillators, using the transformation as $\beta_k = i\sqrt{\frac{\tilde{C}(|\mu_k|)\hslash\omega_k}{2}}\left(\mu_k^*-\mu_k\right)$ and $\alpha_k = \sqrt{\frac{\tilde{C}(|\mu_k|)\hslash}{2\omega_k}}\left(\mu_k^*+\mu_k\right)$.
Here $\tilde{C}(\vert\mu\vert)$ is defined as $\frac{2S}{1+\vert\mu\vert^2}$
Similarly for the system oscillator co-ordinate ${r}$ and momentum ${\rho}$ can be expressed as $r = \sqrt{\frac{\tilde{C}(|\xi|)\hslash}{2\omega_0}}\left(\xi^*+\xi\right)$ and $\rho = i\sqrt{\frac{\tilde{C}(|\xi|)\hslash\omega_0}{2}}\left(\xi^*-\xi\right)$.
Then the Hamiltonian in the coherent state variables can be expressed in terms of these redefined quantities as,
\begin{eqnarray} \label{e2:6}
H = \frac{\rho^2}{2}+\frac{1}{2}\omega_0^2r^2+\sum_k \frac{\beta_k^2}{2} + \frac{1}{2}\sum_k\omega_k^2\left(\alpha_k -\frac{c_k}{\omega_k^2}r \right)^2
\end{eqnarray}
where $c_k$ is given by $2\sqrt{g_kg'_k\omega_k\omega_0\tilde{c}(|\xi|)}$ or $2\sqrt{\frac{\tilde{c}(|\xi|)}{\tilde{c}(|\mu_k|)}}g'_k\sqrt{\omega_k\omega_0}$ after discarding the irrelevant constant terms.
The above Hamiltonian is different from our starting Hamiltonian operator equation \eqref{e2:1} because of the c-number nature of the coherent state variables.
It is thus possible to reduce the quantum dynamics of a spin-$ \frac{1}{2} $ particle in a sea of spin-bath to the dynamics of a particle in an oscillator bath.
This reduction is realizable in view of the well known kinship between the spin-$ \frac{1}{2} $ algebra and the algebra of bosons according to Schwinger's prescription\cite{schwinger}.
Here, we use the Radcliffe coherent states, where Holstein-Primakoff transformation plays a connection between the spin-$ \frac{1}{2} $ algebra and boson operators as does the Schwinger mapping\cite{schwinger}.
Furthermore, high spin limit of a spin coherent state merges to the harmonic oscillator coherent state\cite{radcliffe}.
The spin-bath as a set of oscillators had been realized explicitly earlier\cite{forsythe} in a different context.
The Hamiltonian \eqref{e2:6} is therefore the c-number equivalent of the Hamiltonian operator \eqref{e2:1}.
Thermalization of the spin in a spin bath can thus be conveniently understood in terms of this c-number Hamiltonian.
Although we have formulated the problem for arbitrary length of bath spin S, it is not possible to see the interpolation of the free energy behaviour between spin-$\frac{1}{2}$ (S=$\frac{1}{2}$) and harmonic oscillators (S=$\infty$) limits.
This is because of the fact the calculation requires imposition of statistics which is applicable only for the two limits, \textit{i.e.} Fermi-Dirac for spin-$\frac{1}{2}$ particle and Bose-Einstein for the harmonic oscillator systems.
In the interpolation regime it is not clear, a priori, which statistics the system is going to follow.

The equations of motion for the particle and for the bath variables according to the Hamiltonian Eq. \eqref{e2:6} take the form:
\begin{eqnarray}
\ddot{r}+\omega_0^2r &=& \sum_kc_k\left(\alpha_k-\frac{c_k}{\omega_k^2}r\right) \label{e2:7}\\
\ddot{\alpha}_k+\omega_k^2\alpha_k &=& c_kr \label{e2:8}
\end{eqnarray}
Solving Eq. \eqref{e2:8} and substituting in Eq. \eqref{e2:7} with the condition $r(0) = 0$, we eliminate the bath degrees of freedom to obtain our desired c-number Langevin equation for the particle,
\begin{equation} \label{e2:9}
\ddot{r}+\int_0^t dt'\gamma(t-t')\dot{r}(t')+\omega_0^2r = \eta(t)
\end{equation}
where,
\begin{eqnarray}
\gamma(t)&=&\sum_k\frac{c_k^2}{\omega_k^2}\;\cos \omega_kt \label{e2:10}\\ \eta(t)&=&\sum_kc_k\alpha_k(0)\cos\omega_kt+\frac{c_k}{\omega_k}\beta_k(0)\sin\omega_kt \label{e2:11}
\end{eqnarray}
are the memory kernel and noise, respectively.

In order to quantify the properties of the thermal bath, it is convenient to introduce a spectral density function $J(\omega)$ associated with the system-bath interaction.

\begin{equation} \label{e2:12}
J(\omega) = \frac{\pi}{2}\sum_k\frac{c_k^2}{\omega_k}\delta(\omega-\omega_k)
\end{equation}

In terms of spectral density function $J(\omega)$, one may rewrite the expressions for memory kernel in Eq. \eqref{e2:10} as,
\begin{eqnarray}
\gamma(t)=\frac{2}{\pi}\int_{-\infty}^\infty d\omega\frac{J(\omega)}{\omega}\;\cos \omega t \label{e2:13}
\end{eqnarray}
while the noise $\eta(t)$ must satisfy the characteristics of the spin bath at equilibrium:
\begin{eqnarray}
\left\langle\eta(t)\right\rangle_s&=& 0 \label{e2:14}\\
\left\langle\eta(t)\eta(t')\right\rangle_s&=&\frac{2}{\pi}\int_{-\infty}^\infty d\omega\frac{J(\omega)}{\omega}\left[\frac{\hslash\omega}{2}tanh(\frac{\hslash\omega}{2KT})\right]\cos \omega(t-t') \label{e2:15}
\end{eqnarray}

To ensure that the c-number noise $\eta(t)$ is zero-centred (Eq. \eqref{e2:14}) and satisfies fluctuation-dissipation relation (Eq. \eqref{e2:15}), it is necessary to assume a canonical distribution of Gaussian form for statistical averaging $\left\langle ... \right\rangle_s$ over c-number bath variables as follows:
\begin{equation} \label{e2:16}
P_k(\alpha_k(0),\beta_k(0)) = N exp\left(-\frac{\frac{1}{2}\beta_k^2(0)+\frac{1}{2}\omega_k^2\alpha_k^2(0)}{2tanh(\frac{\hslash\omega_k}{2KT})}\right)
\end{equation}
This is the spin bath counterpart\cite{sinha_pre,ghosh_jcp} of Wigner canonical thermal distribution function\cite{hillery} for the harmonic or bosonic bath.
Here N is the normalization constant.
The width of distribution is given by $ tanh(\frac{\hslash\omega_k}{2KT}) $, which is related to the average thermal excitation number $\bar{n}_F(\omega_k)$ of the bath as $tanh(\frac{\hslash\omega_k}{2KT}) = 1-2\bar{n}_F(\omega_k)$, $\bar{n}_F(\omega_k)$ being Fermi-Dirac distribution function.
The width of Wigner distribution on the other hand is determined by $ coth(\frac{\hslash\omega_k}{2KT}) $.
Note that at T=0, both distributions merge at a single value.
The differences begin to appear at finite temperatures.
At high temperatures, coth-factor reduces to a factor that results in the recovery of the classical limit.
On the other hand the distribution (Eq.\eqref{e2:16}) for the spin bath does not.
Therefore the thermalization of the particle in a spin-bath can be described by this canonical thermal distribution \eqref{e2:16} and the c-number Hamiltonian (Eq.\ref{e2:6}).

\section{Thermal equilibrium of a spin coupled to a spin-bath}
\subsection{General expression for specific heat}
Here we derive the expression for specific heat ($ C_v $) of a spin-spin-bath system using normal mode frequencies and examine the high and low temperature limits.
We expect from the discussions in the previous section, that at low temperatures the spin-bath closely follows a bosonic bath while it differs at high temperatures.

We begin by considering the normal mode solutions of Eqs. \eqref{e2:7} and \eqref{e2:8}.
To this end we write,
\begin{eqnarray} \label{e3:1}
r(t)&=& r_0(\omega) e^{-i\omega t} \nonumber\\
\alpha_k(t)&=& \alpha_k(\omega) e^{-i\omega t}\;\;\;k=1,2,...
\end{eqnarray}
Using the solutions \eqref{e3:1} in Eqs \eqref{e2:7} and \eqref{e2:8} we obtain,
\begin{eqnarray}
\left(\omega_0^2-\omega^2\right)r_0(\omega)&=& \sum_k c_k\left(\alpha_k(\omega)-\frac{c_k}{\omega_k^2}r_0(\omega)\right) \label{e3:2}\\
\left(\omega_k^2-\omega^2\right)\alpha_k(\omega)&=& c_k\;r_0(\omega) \label{e3:3}
\end{eqnarray}
Elimination of $r_0(\omega)$ from Eqs. \eqref{e3:2} and \eqref{e3:3} yields,
\begin{equation} \label{e3:4}
\left(\omega^2-\omega_0^2\right)= \sum_k\frac{c_k^2}{\omega_k^2}\left(\frac{\omega^2}{\omega^2-\omega_k^2}\right)
\end{equation}
From the above equation the generalised susceptibility can be defined\cite{ullersma_pa} as,
\begin{equation} \label{e3:5}
\kappa(\omega)=\left[-\left(\omega^2-\omega_0^2\right)+\sum_k\frac{c_k^2}{\omega_k^2}\left(\frac{\omega^2}{\omega^2-\omega_k^2}\right)\right]^{-1}
\end{equation}
For discrete modes, $\kappa(\omega)$ has poles on the real axis at the normal mode frequencies of the system-plus-bath and has zeros at normal mode frequencies of the bath only.
This allows one to write $\kappa(\omega)$ in terms of the products of the ratios of the functions of the appropriate normal mode frequencies $\Pi_i(\omega^2-\omega_i^2)/\Pi_j(\omega^2-\omega_j^2)$, $ i $ and $ j $ indices correspond to bath and system-plus-bath normal modes, respectively.
Using Eqs.\eqref{e2:10} and \eqref{e2:12}, the above equation can be rewritten;
\begin{equation} \label{e3:6}
\kappa(\omega)=\left[-\omega^2+\omega_0^2-i\omega\tilde{\gamma}(\omega)\right]^{-1}
\end{equation}
where $\tilde{\gamma}(\omega)$ is Laplace transform of $\gamma(t)$.
In what follows we show that $\kappa(\omega)$ is responsible for dynamical corrections.
We now consider that the spin coupled to the spin bath is in thermal equilibrium at temperature T.
This system has well defined free energy F(T) which can be expressed\cite{ford_jsp} as the difference between the free energy of spin-spin bath system and the free energy of the bath in absence of the spin.
This can be related to the dynamical susceptibility $\kappa(\omega)$ as follows;
\begin{equation} \label{e3:7}
F=\frac{1}{\pi}\int_0^\infty d\omega f(\omega,T)Im\left[\frac{d}{d\omega}ln|\kappa(\omega)|\right]
\end{equation}
where
\begin{equation} \label{e3:8}
f(\omega,T)=-KT\,ln\left[1+exp\left(-\frac{\hslash\omega}{KT}\right)\right]
\end{equation}
is the free energy of a spin-$ \frac{1}{2} $ particle or two-level atom at frequency $\omega$.
The thermodynamic functions, entropy and specific heat can be derived from the following relations:
\begin{eqnarray}
S &=& -\frac{\partial F}{\partial T}\label{e3:9}\\
C_v &=& T\frac{\partial S}{\partial T}=-T\frac{\partial^2 F}{\partial T^2} \label{e3:10}
\end{eqnarray}

Now considering to heat bath to be Ohmic, i.e., $\tilde{\gamma}(\omega)=\gamma_0$ one can simplify Eq. \eqref{e3:6} as follows,
\begin{equation} \label{e3:11}
\kappa(\omega)=\left[-\omega^2+\omega_0^2-i\omega\gamma_0\right]^{-1}
\end{equation}
which results in
\begin{eqnarray} \label{e3:12}
Im\left[\frac{d}{d\omega}\left( ln\vert\kappa(\omega)\vert\right) \right]=\frac{\gamma_0\left(\omega^2+\omega_0^2\right)} {\left(\omega^2-\omega_0^2\right)^2+\gamma_0^2\omega^2}
\end{eqnarray}
Then Eq.\eqref{e3:7} becomes,
\begin{eqnarray} \label{e3:13}
F(T)&=& \frac{1}{\pi}\int_0^\infty d\omega f(\omega, T)\left[ \frac{\gamma_0\left(\omega^2+\omega_0^2\right)} {\left(\omega^2-\omega_0^2\right)^2+\gamma_0^2\omega^2}\right] \nonumber\\
&=& \frac{1}{\pi}\int_0^\infty d\omega f(\omega, T)\left( \frac{z}{\omega^2+z^2}+\frac{z^*}{\omega^2+{z^*}^{2}}\right)
\end{eqnarray}
where, $z=\frac{\gamma_0}{2}+i\omega_1$ and $z^*=\frac{\gamma_0}{2}-i\omega_1$ and $\omega_1=\sqrt{\omega_0^2-\frac{\gamma_0^2}{4}}$.
Eq. \eqref{e3:13} can be rewritten as,
\begin{eqnarray} \label{e3:14}
F(T)&=& \frac{1}{\pi}\int_0^\infty d\omega f(\omega, T)\left( \frac{z}{\omega^2+z^2}+\frac{z^*}{\omega^2+z^{*2}}\right) \nonumber\\
&=& \bar{g}(z) + \bar{g}(z^*)
\end{eqnarray}
Here $ \bar{g}(z) $ can be expressed using substitution $\frac{\hslash\omega}{KT}=x$.
\begin{eqnarray} \label{e3:15}
\bar{g}(z)&=& -\frac{KT}{\pi}\int_0^\infty dx\; ln(1+e^{-x}) \frac{\frac{\hslash z}{KT}}{\left(\frac{\hslash z}{KT}\right)^2+x^2}
\end{eqnarray}
Furthermore defining $\frac{\hslash z}{KT} = y$ we obtain,
\begin{eqnarray} \label{e3:16}
g(y)&=& -\frac{KT}{\pi}\int_0^\infty dx\; ln(1+e^{-x}) \frac{y}{y^2+x^2}\nonumber\\
and \qquad\qquad\qquad g(y^*)&=& -\frac{KT}{\pi}\int_0^\infty dx\; ln(1+e^{-x}) \frac{y^*}{y^{*2}+x^2}
\end{eqnarray}
Therefore, we can write down Eq.\eqref{e3:14} as
\begin{eqnarray} \label{e3:17}
F(y)&=& g(y)+g(y^*)\nonumber\\
&=&-\frac{KT}{\pi}\int_0^\infty dx\; ln(1+e^{-x})\left[ \frac{y}{y^2+x^2}+\frac{y^*}{y^{*2}+x^2}\right]
\end{eqnarray}
This is the central result of this paper which is valid for arbitrary temperature.
We now consider two different limits (low and high temperature) depending upon the parameter value $|y|$.

\subsection{Temperature dependence of Specific heat}
\subsubsection{Low-Temperature limit:($ |y|\gg 1, KT\ll\hslash|z| $)}
Partial integration of Eq.\eqref{e3:16} results in,
\begin{eqnarray} \label{e3:18}
g(y)&=& -\frac{KT}{\pi}\int_0^\infty dx \frac{tan^{-1}(\frac{x}{y})}{e^x+1}
\end{eqnarray}

When the parameter $|y|\gg 1$, corresponding to $KT\ll\hslash |z|$ (which, in turn, implies $KT\ll\hslash\omega_0$), the argument of $tan^{-1}(\frac{x}{y})$ can be expanded as,
\begin{equation} \label{e3:19}
tan^{-1}\left(\frac{x}{y}\right) = \sum_{n=0}^{\infty}\frac{(-1)^n}{2n+1}\left(\frac{x}{y}\right)^{2n+1}
\qquad \qquad\left|\frac{x}{y}\right| \ll 1
\end{equation}
We then obtain,
\begin{eqnarray} \label{e3:20}
g(y)&=& \frac{KT}{\pi}\sum_{n=0}^{\infty}\frac{(-1)^{n+1}}{(2n+1)y^{2n+1}}\int_0^\infty dx \frac{x^{2n+1}}{e^x+1}\nonumber\\
&=& \frac{KT}{\pi}\sum_{n=0}^{\infty}\frac{(-1)^{n+1}}{(2n+1)y^{2n+1}}\zeta(2n+2)\Gamma(2n+2)\left(1-\frac{1}{2^{2n+1}}\right)
\end{eqnarray}
where $\zeta(n)$ is Reimann-zeta function and $\Gamma(n)$ is Gamma function.
Now using the above results and substituting the values of $g(y)$ and $g(y^*)$ in Eq.\eqref{e3:17}, we have
\begin{eqnarray} \label{e3:21}
F(T)&=&\frac{1}{\pi}\sum_{n=0}^{\infty}\frac{(-1)^{n+1}(KT)^{2n+2}}{(2n+1)\hslash^{2n+1}}\zeta(2n+2)\Gamma(2n+2)\left(1-\frac{1}{2^{2n+1}}\right)\left(\frac{z^{2n+1}+z^{*(2n+1)}}{|z|^{4n+2}}\right)\nonumber\\
&=& \frac{1}{\pi}\sum_{n=0}^{\infty}\frac{(-1)^{n+1}(KT)^{2n+2}}{(2n+1)\hslash^{2n+1}}\zeta(2n+2)\Gamma(2n+2)\left(1-\frac{1}{2^{2n+1}}\right)\frac{1}{|z|^{4n+2}}\nonumber\\
& &\left(\left(z+z^*\right)^{2n+1}-\sum_{k=1}^n \left(\begin{array}{c}2n+1 \\ k \\\end{array}\right)|z|^{2k}\left(z^{2(n-k)+1}+z^{*(2(n-k)+1)}\right)\right)
\end{eqnarray}
The expression for $C_v$ follows immediately,
\begin{eqnarray} \label{e3:22}
C_v&=& -T\frac{\partial^2F}{\partial T^2} \nonumber\\
&=& \frac{K}{\pi}\sum_{n=0}^\infty\frac{(-1)^n(2n+2)}{|z|^{4n+2}}\left(\frac{KT}{\hslash}\right)^{2n+1}\zeta(2n+2)\Gamma(2n+2)\left(1-\frac{1}{2^{2n+1}}\right)\nonumber\\
& &\left[\left(z+z^*\right)^{2n+1}-\sum_{k=1}^n\left(\begin{array}{c}2n+1 \\ k \\\end{array}\right)|z|^{2k}\left(z^{2(n-k)+1}+z^{*(2(n-k)+1)}\right)\right]
\end{eqnarray}
From Eq.\eqref{e3:22} the specific heat up to three leading order terms are,
\begin{eqnarray} \label{e3:23}
C_v&=& K\left[\left\lbrace\frac{\pi}{6}\left(\frac{KT}{\hslash\omega_0}\right)\right\rbrace\left\lbrace\frac{\gamma_0}{\omega_0}\right\rbrace+\left\lbrace\frac{7\pi^3}{30}\left(\frac{KT}{\hslash\omega_0}\right)^3\right\rbrace\left\lbrace 3\left(\frac{\gamma_0}{\omega_0}\right)-\left(\frac{\gamma_0}{\omega_0}\right)^3\right\rbrace\nonumber\right.\\
& &\left.+\left\lbrace\frac{31\pi^5}{42}\left(\frac{KT}{\hslash\omega_0}\right)^5\right\rbrace\left\lbrace \left(\frac{\gamma_0}{\omega_0}\right)^5-5\left(\frac{\gamma_0}{\omega_0}\right)^3+5\left(\frac{\gamma_0}{\omega_0}\right)\right\rbrace\right]
\end{eqnarray}
A close inspection of the above expression clearly reveals that each term in the square bracket is a product of two terms.
The first one is temperature dependent and of the form $\left(\frac{KT}{\hslash\omega_0}\right)^n$, with n=1,3,5 and originates from traditional thermodynamics.
The second one contains powers of $\left(\frac{\gamma_0}{\omega_0}\right)$ which has a dynamical origin in $ \kappa(\omega) $ and is the major content of this work.
The appearance of dynamical factor is reminiscent of the Kramers theory of activated rate process where the rate constant from transition state theory (thermodynamic contribution) gets multiplied by this factor in the form $\left(\frac{\omega_b}{\gamma_0}\right)$ where $\omega_b$ is the frequency of the inverted oscillator well at the barrier top.
Drawing a hint from this observation, we may therefore emphasize that the above expression for $C_v$ at low temperature pertains to a small system allowing a steady flow of energy from it to the environment in the form of dissipation under a quasi-stationary condition.
Secondly, we note that the expression for specific heat matches with that for the well known form for degenerate fermi system, i.e., $C_v=AT+BT^3$, (A, B being constants), so far as the thermal behaviour of the system is concerned.
Our results qualitatively agree with the results obtained by Ford and O'connell\cite{ford_prb} and by H\"anggi and Ingold\cite{hanggi_app} for a harmonic oscillator in a harmonic bath.
The origin of this agreement lies on merging of the thermal behaviour of the spin bath and the harmonic bath as T $\rightarrow$ 0 as discussed earlier.
The specific characteristics of the spin system are, however, reflected in the details of numerical factors.

\subsubsection{High-Temperature limit:($ |y|\rightarrow 0, KT\gg \hslash\omega_0 $)}
In order to calculate the temperature dependence in the high temperature limit, we expand the logarithmic term in Eq.\eqref{e3:16} as,
\begin{equation} \label{e3:24}
ln(1+e^{-x}) = \sum_{n=1}^{\infty}(-1)^n\frac{e^{-nx}}{n}\qquad\qquad\left[e^{-x}\ll 1\right]
\end{equation}
Substituting Eq.\eqref{e3:24} in Eq. \eqref{e3:16} we obtain,
\begin{eqnarray} \label{e3:25}
g(y)&=& \frac{KT}{\pi}\sum_{n=1}^{\infty}(-1)^{n-1}\int_0^\infty dx\frac{y}{y^2+x^2}\frac{e^{-nx}}{n} \nonumber\\
&=& \frac{KT}{\pi}\sum_{n=1}^{\infty}(-1)^{n-1}\left[Ci(ny)Sin(ny)+\frac{1}{2} Cos(ny)(\pi-2Si(ny))\right]
\end{eqnarray}
Making use of Eqs\eqref{e3:25} in Eq.\eqref{e3:16} and after formally expanding upto fourth order of $y$ we obtain,
\begin{eqnarray} \label{e3:26}
g(y)&=& \frac{K T}{\pi}\left[(\gamma_e-2)y+y^2\sum_{n=1}^\infty(-1)^{n-1}n-\left(\gamma_e-\frac{4}{3}\right)\sum_{n=1}^{\infty}(-1)^{n-1}n^2\frac{y^3}{6}\right.\nonumber\\
& & \left.+\sum_{n=1}^{\infty}(-1)^{n-1}n^3\frac{y^4}{6}+\frac{\pi}{2}\left(1-\sum_{n=1}^{\infty}(-1)^{n-1}n\frac{y^2}{2!}+\sum_{n=1}^{\infty}(-1)^{n-1}n^3\frac{y^4}{4!}\right)\right]
\end{eqnarray}
where $\gamma_e = 0.577216$.
After substitution of the values of $y, z, z^*$, the free energy in the high temperature regime is given by,
\begin{eqnarray} \label{e3:27}
F(T) &=& \frac{KT}{\pi}\left[(\gamma_e-2)\hslash\gamma_0+\sum_{n=1}^{\infty}\frac{(-1)^{n-1}}{n}\left(\frac{n\hbar\omega_0}{KT}\right)^2\left\lbrace\frac{4-\pi}{4}\left(\left(\frac{\gamma_0}{\omega_0}\right)^2-2\right)\right\rbrace\right.\nonumber\\
&-& \left.\sum_{n=1}^{\infty}\frac{(-1)^{n-1}}{n}\left(\frac{n\hbar\omega_0}{KT}\right)^3\left\lbrace\frac{\gamma_e-4/3}{6}\left(\left(\frac{\gamma_0}{\omega_0}\right)^3-3\left(\frac{\gamma_0}{\omega_0}\right)\right)\right\rbrace+\frac{\pi}{2}KT\right. \nonumber\\
&+&\left.\sum_{n=1}^{\infty}\frac{(-1)^{n-1}}{n}\left(\frac{n\hbar\omega_0}{KT}\right)^4\left\lbrace\frac{8+\pi}{48}\left(\left(\frac{\gamma_0}{\omega_0}\right)^4+2-4\left(\frac{\gamma_0}{\omega_0}\right)^2\right)\right\rbrace\right]
\end{eqnarray}
From Eq.\eqref{e3:27} the expression for specific heat $C_v$ can be calculated as follows;
\begin{eqnarray} \label{e3:28}
C_v&=& -T\frac{\partial^2F}{\partial T^2} \nonumber\\
&=& \frac{K}{\pi}\left[\sum_{n=1}^{\infty}\frac{(-1)^{n}}{n}\left(\frac{n\hbar\omega_0}{KT}\right)^2\left\lbrace\frac{(4-\pi)}{2} \left(\left(\frac{\gamma_0}{\omega_0}\right)^2-2\right)\right\rbrace\right.\nonumber\\
&-& \left.\sum_{n=1}^{\infty}\frac{(-1)^{n}}{n}\left(\frac{n\hbar\omega_0}{KT}\right)^3\left\lbrace(\gamma_e-4/3)\left(\left(\frac{\gamma_0}{\omega_0}\right)^3-3\left(\frac{\gamma_0}{\omega_0}\right)\right)\right\rbrace\right. \nonumber\\
&+&\left.\sum_{n=1}^{\infty}\frac{(-1)^{n}}{n}\left(\frac{n\hbar\omega_0}{KT}\right)^4\left\lbrace\frac{(8+\pi)}{4}\left(\left(\frac{\gamma_0}{\omega_0}\right)^4+2-4\left(\frac{\gamma_0}{\omega_0}\right)^2\right)\right\rbrace\right]
\end{eqnarray}
Eq.\eqref{e3:28} shows that the quantities in the square bracket is again a product of an equilibrium factor and a dynamical factor that depends on the ratio $ \frac{\gamma_0}{\omega_0} $.
In order to extract out the physically relevant form of the above expression we need to consider the thermal saturation effect at high temperature since the retention of higher order terms in Eq.\eqref{e3:28} does not make it meaningful above the saturation temperature.
This is because the system gets decoupled from the bath above this temperature\cite{shao_prl,prokofev,ghosh_jcp}.
This can be ascertained from the effective spectral density and by expressing $\frac{1}{2}tanh\left(\frac{\hslash\omega}{2KT}\right)=\frac{1}{2}-\frac{1}{e^{\hslash\omega/KT}+1} = -\left\langle\hat{S}_z\right\rangle$ where $\left\langle\hat{S}_z\right\rangle$ is a measure of the population difference between the two levels of a bath atom.
We note that as $\left(e^{\hslash\omega/KT} +1\right) \rightarrow 2$ the hyperbolic tangent factor tends to vanish as a result of thermal saturation of the bath.
Keeping therefore only the leading order term of Eq.\eqref{e3:28}, $C_v$ at high temperature can be expressed as,
\begin{eqnarray} \label{e3:29}
C_v = K\left[\left\lbrace\frac{1}{4}\left(\frac{\hslash\omega_0}{KT}\right)^2\right\rbrace\left\lbrace\frac{(8-2\pi)}{\pi} \left(\left(\frac{\gamma_0}{\omega_0}\right)^2-2\right)\right\rbrace\right]
\end{eqnarray}
In order to check this result, we must compare it with the leading order term of specific heat of a single spin system in thermal equilibrium.
The average energy of a single spin is,
\begin{eqnarray} \label{e3:30}
E &=& \hslash\omega_0\left\langle\hat{S}_z\right\rangle
\end{eqnarray}
Substituting $\left\langle\hat{S}_z\right\rangle = -\frac{1}{2}tanh(\frac{\hslash\omega_0}{2KT})$ in Eq.\eqref{e3:30} and expanding in inverse powers of T\cite{das}, we obtain the leading order term;
\begin{eqnarray} \label{e3:31}
E &=& -\frac{(\hslash\omega_0)^2}{4KT} 
\end{eqnarray}
and the specific heat as,
\begin{eqnarray} \label{e3:32}
C_v &=& K\frac{1}{4}\left(\frac{\hslash\omega_0}{KT}\right)^2 
\end{eqnarray}
This corresponds to the equilibrium factor in the expression for specific heat at high temperature in Eq.\eqref{e3:29} and serves as an important check for our calculation.

\begin{figure}[!ht]
\includegraphics[scale=1,angle=0,width=10cm]{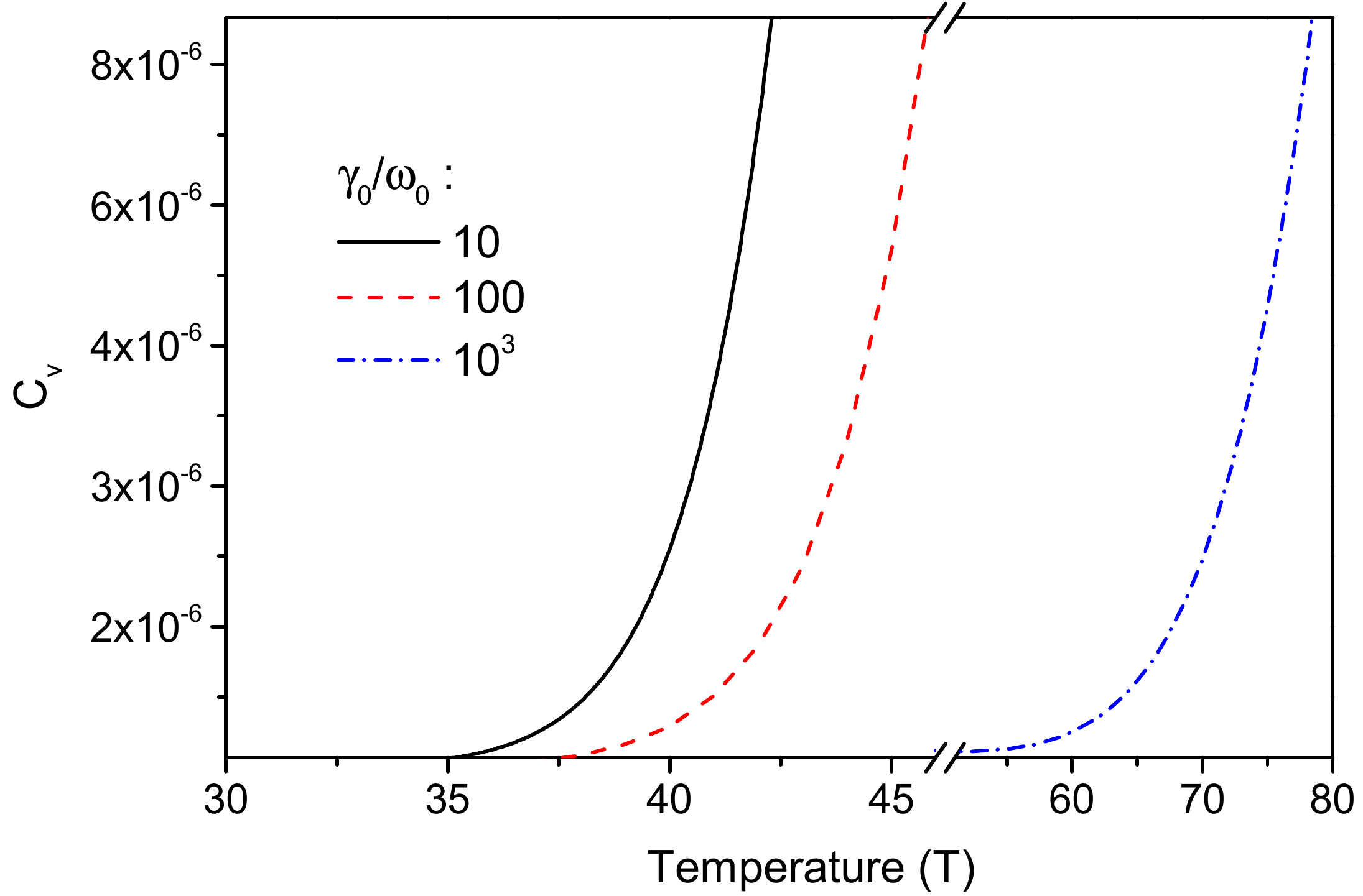}
\caption{Variation of specific heat with temperature for three values of the ratio $\gamma_0/\omega_0$ calculated using Eq.\eqref{e3:13} in the low temperature regime. (scale arbitrary)} \label{f1}
\end{figure}

\begin{figure}[!ht]
\includegraphics[scale=1,angle=0,width=10cm]{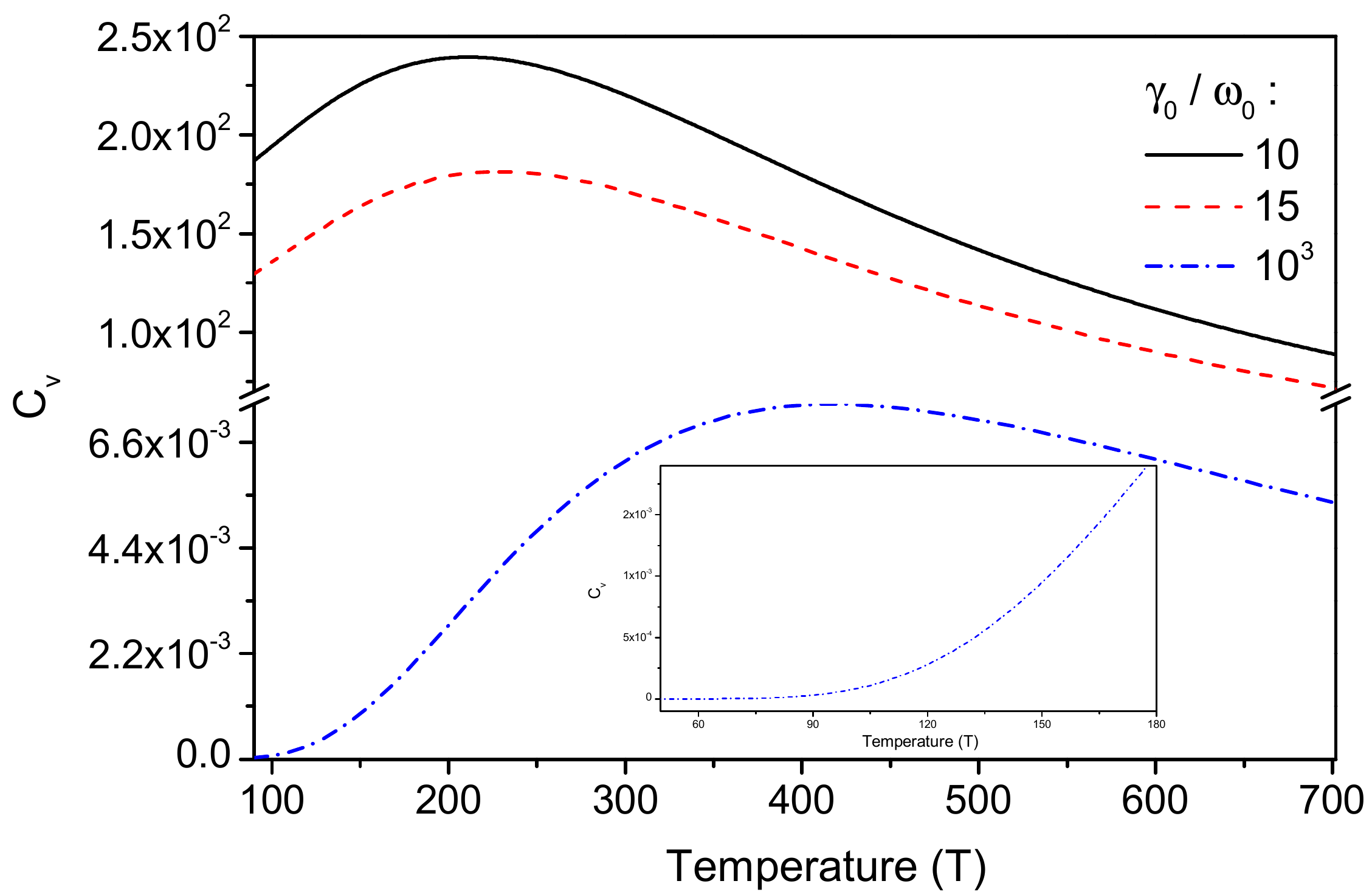}
\caption{The same as in Fig.1 but for the high temperature regime. (Inset)Plot of specific heat shows smooth crossover from low to high temperature regime for $\gamma_0:\omega_0 = 10^3$. (scale arbitrary)} \label{f2}
\end{figure}

\begin{figure}[!ht]
\includegraphics[scale=1,angle=0,width=10cm]{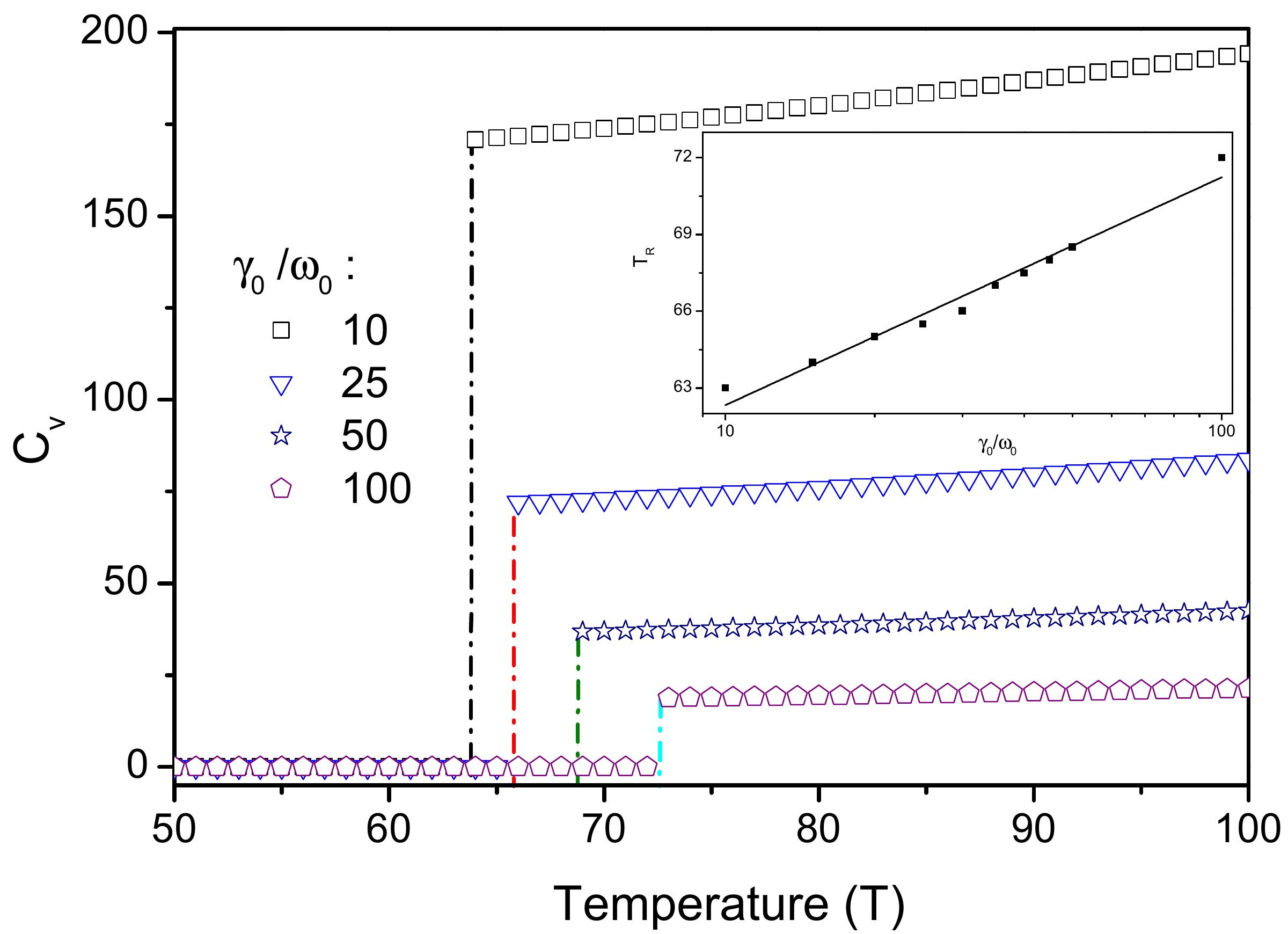}
\caption{Transition of specific heat for several values of $\gamma_0/\omega_0$. (Inset) Variation of transition temperature $T_R$ vs $\gamma_0/\omega_0$ in a semilog plot. (scale arbitrary)} \label{f3}
\end{figure}

\subsection{Dynamical factors and specific heat; finite size effect}
We now examine the variation of specific heat with temperature for several values of dissipation parameters.
The expression for free energy in an integral form Eq.\eqref{e3:17} is evaluated numerically to obtain the specific heat over a wide range of temperature.
The results are shown in Figs. 1-3.
Fig. 1 depicts the variation in the low temperature regime which follows $AT+BT^3$ law, where $A$ and $B$ are constants.
The behaviour in the high temperature regime is illustrated in Fig.2.
The specific heat follows characteristically $C_v\sim T^{-2}$ behaviour.
It is further evident that the variation of specific heat is accompanied by a sharp transition as one passes from low to high temperature regime.
This transition is illustrated in Fig.3.
For smaller values of dissipation constant, i.e. when the system size is small, the transition is marked by a larger jump between the two temperature regimes and a shift towards the lower temperature.
For larger values of dissipation constant one observes a comparatively smooth crossover of specific heat between the two temperature regions as shown in inset of Fig.2.
The variation of transition temperature as a function of the ratio of the dissipation constant and characteristic frequency satisfies a linear fit on a semi-logarithmic scale as shown in inset of Fig.3.
We mention, in passing, that the observed correlation between the dynamical factors and the specific heat for finite size systems are reminiscent of the recent experimental results \cite{tang} on the behaviour of specific heat with temperature in cobalt oxide layers and micrograins, characterized by anti-ferromagnetic transition.
One observes significant reduction of magnetic ordering temperature or Neel temperature for thin layers, \textit{i.e.} for smaller system size.
We however, emphasize that this correlation should not be extended further for a quantitative comparison since in our model the spin bath consists of independent spins, whereas the experimental situation corresponds to interacting spins via exchange interaction of order of Neel temperature.
A more suitable experimental systems for probing the correlation between dynamical factors and the specific heat may be the two-level system in super-conducting circuits coupled to a fermionic bath\cite{paladino} employed for the study of decoherence.

\section{Conclusion}\label{s6}
In this work we have examined the behaviour of specific heat of a spin-$\frac{1}{2}$ system coupled to a spin-bath with finite coupling strength.
The basis of our analysis is the mapping of the spin-spin-bath Hamiltonian into its oscillator counterpart using Holstein-Primakoff transformation followed by subsequent diagonalization of the Hamiltonian in its coherent state representation.
The derived expression for the specific heat is characterized by the dynamical correction factors over and above the usual thermal equilibrium factors.
By tuning the system size it is possible to realize this dynamical contribution to specific heat and its consequences.
We summarize the main conclusions of this study as follows:

(i) The c-number quantum Langevin equation assumes exactly the traditional classical looking form, the essential difference, however, lies on the nature of noise characteristics due to the spin-bath.
Therefore the susceptibility, the key quantity for calculation of free energy associated with the quantum Langevin equation retains also its standard form.
The underlying universality of the spin bath is therefore reflected in the analysis so far as the dynamic behaviour is concerned.

(ii) The expression for specific heat both in the high and low temperature regions appears as a sum of the products of a thermal equilibration factor and a dynamical correction factor the later being a function of the ratio of the friction coefficient and the characteristic frequency of the system.
Up to a leading order the specific heat at low temperature can be expressed as $C_v=A(\gamma_0/\omega_0)T+B(\gamma_0/\omega_0)T^3$ which reflects the general behaviour of the degenerate fermi system.
The modification by dissipative contributions due to Ohmic bath is very much similar to that for the bosonic bath.
This again highlights the similarities of the behaviour of the spin bath and the harmonic bath as $T\rightarrow 0$.

(iii) The high temperature behaviour of the spin bath is dominated by thermal saturation of the two levels of the spin system.
Up to a leading order the specific heat below saturation temperature can be expressed as $C_v=C(\gamma_0/\omega_0)\frac{1}{T^2}$.
The result corresponds to the qualitative behaviour of a single spin-$ \frac{1}{2} $ system.
The transition between the low and high temperature regimes is marked by abrupt jump accompanied by shift towards lower temperature for smaller dissipation characteristic of the reduced system size.
For higher dissipation \textit{i.e.}, in the thermodynamic limit, we observe a smooth crossover of specific heat.

\section{Acknowledgement}\label{s7}
Thanks are due to the Council of Scientific and Industrial Research, Government of India, for partial financial support.


\begin{center}
References
\end{center}
\begin{thebibliography}{references} \label{ref}
%
\bibitem{hanggi_rmp}
R. Zwanzig, J. Stat. Phys. 9, 215 (1973); P. H\"anggi, P. Talkner, and M. Borkovec, Rev. Mod. Phys. \textbf{62}, 251 (1990); A. J. Leggett, S. Chakravarty, A. T. Dorsy, M. P. A. Fisher, A. Garg and W. Zwerger, Rev. Mod. Phys. \textbf{67}, 725(E) (1995), ibid. \textbf{59}, 1 (1987).
%
\bibitem{grabert}
H. Grabert, P. Schramm, and G. L. Ingold, Phys. Rep. 168, 115 (1988); A. J. Leggett, S. Chakravarty, A. T. Dorsey, M. P. A. Fisher, A. Garg, and W. Zwerger, Rev. Mod. Phys. 59, 1 (1987); ibid. 67, 725(E) (1995).
%
\bibitem{louisell}
W. H. Louisell, \textit{Quantum Statistical Properties of Radiation} (John Wiley, New York, 1973); P. Meystre and M. Sargent, \textit{Elements of Quantum Optics} (Springer-Verlag, New York, 1991); U. Weiss, \textit{Quantum Dissipative Systems} (World Scientific, Singapore, 1999).
%
\bibitem{nitzan}
A. Nitzan, \textit{Chemical Dynamics in Condensed Phase} (Oxford University Press, New York, 2006); V. May and O. K\"uhn, \textit{Charge and Energy Transfer Dynamics in Molecular Systems} (Wiley-VCH, Weinheim, 2004); D. Barik, D. Banerjee, and D. S. Ray, \textit{Quantum Brownian Motion in c-Numbers: Theory and Application} (Nova-Science Publishers, New York, 2009).
%
\bibitem{ford_prl} 
G. W. Ford, J. T. Lewis and R. F. O'connell, Phys. Rev. Lett, \textbf{55}, 2273 (1985).
%
\bibitem{hanggi_njp}
P. H\"anggi, G. L. Ingold and P. Talkner, New J. Phys., \textbf{10}, 115008 (2008).
%
\bibitem{ford_prb}
G. W. Ford and R. F. O'connell, Phys. Rev. B, \textbf{75}, 134301 (2007).
%
\bibitem{hanggi_app}
P. H\"anggi and G. L. Ingold, Acta Phys. Polo., \textbf{37}, 1537 (2006).
%
\bibitem{ferrer}
A. V. Ferrer and C. M. Smith, Phys. Rev. B \textbf{76}, 214303 (2007); A. O. Caldeira, A. H. Castro Neto, and T. Oliveira de Carvalho Phys. Rev. B \textbf{48}, 13974 (1993); A. V. Ferrer, A. O. Caldeira, and C. M. Smith, Phys. Rev. B \textbf{74}, 184304 (2006).
%
\bibitem{tang}
Y. J. Tang, David J. Smith, B. L. Zink, F. Hellman, and A. E. Berkowitz, Phys. Rev. B, \textbf{67}, 054408 (2003).
%
\bibitem{kodama_prb}%
R. H. Kodama and A. E. Berkowitz, Phys. Rev. B 59, 6321 (1999); N. Abarra, K. Takano, F. Hellman, and A. E. Berkowitz, Phys. Rev. Lett. 77, 3451 (1996).
%
\bibitem{viswanatha}%
R. Viswanatha , S. Sapra , B. Satpati , P. V. Satyam , B. N. Dev and D. D. Sarma, J. Mater. Chem., \textbf{14}, 661 (2004).
%
\bibitem{stichtenoth}
D. Stichtenoth, C. Ronning, T. Niermann, L. Wischmeier, T. Voss, C. J. Chien, P. C. Chang and J. G. Lu, Nanotechnology, \textbf{18}, 435701 (2007); A. Wood, M. Giersig, M. Hilgendorff, A. V. Campos, L. M. L. Marz$ \acute{a} $n and P. Mulvaney, Aust. J. Chem. \textbf{56}, 1051 (2003);
%
\bibitem{sinha_pre}%
S. S. Sinha, D. Mondal, B. C. Bag, and D. S. Ray, Phys. Rev. E \textbf{82}, 051125 (2010); S. S. Sinha, A. Ghosh, and D. S. Ray, \textbf{84}, 031118 (2011); \textbf{84}, 041113 (2011).
%
\bibitem{ghosh_jcp}
A. Ghosh, S. S. Sinha, D. S. Ray, J. Chem. Phys., \textbf{134}, 094114 (2011); Phys. Rev. E \textbf{83}, 061154 (2011).
%
\bibitem{ford_jsp}%
G. W. Ford, J. T. Lewis and R. F. O'Connell, J. Stat. Phys., \textbf{53}, 439 (1988).
%
\bibitem{radcliffe}%
J. M. Radcliffe, J Phys A:Math.Gen, \textbf{4}, 313 (1971).
%
\bibitem{holstein}%
T. Holstein and H. Primakoff, Phys. Rev. \textbf{58}, 1098 (1940).
%
\bibitem{ghosh_pre}%
A. Ghosh, S. S. Sinha, D. S. Ray, Phys. Rev. E \textbf{86}, 011122 (2012).
%
\bibitem{klauder}
J. R. Klauder and B. S. Skagerstam, Coherent States, Applications in Physics and Mathematical Physics (World Scientific, Singapore, 1985).
%
\bibitem{schwinger}
J. Schwinger, \textit{Quantum theory of angular momentum Academic Press}, (New York, 1965).
%
\bibitem{forsythe}%
K. M. Forsythe and N. Makri, Phys. Rev. B \textbf{60}, 972 (1999).
%
\bibitem{hillery}
M. Hillery, R. F. O'Connell, M. O. Scully and E. P. Wigner, Phys. Rep., \textbf{106}, 121 (1984).
%
\bibitem{ullersma_pa}
P. Ullersma, Physica \textbf{32}, 27 (1966).
%
\bibitem{shao_prl}
J. Shao and P. H\"anggi, Phys. Rev. Lett. \textbf{81}, 5710 (1998).
%
\bibitem{prokofev}%
N. V. Prokof'ev and P. C. E. Stamp, Rep. Prog. Phys. \textbf{63}, 669 (2000); Phys. Rev. Lett. \textbf{80}, 5794, (1998); J. Low Temp. Phys. \textbf{104}, 143 (1996).
%
\bibitem{das}
A. Das, \textit{Field Theory: A Path Integral Approach}, World Scientific, Singapore, 1993.
%
\bibitem{molina_prb}
M. Molina-Ruiz, A. F. Lopeandia, F. Pi, D. Givord, O. Bourgeois, and J. Rodriguez-Viejo, Phys. Rev. B \textbf{83}, 140407 (2011).
%
\bibitem{binder_phy}
K. Binder, Physica, \textbf{62}, 508 (1972).
%
\bibitem{paladino}
E. Paladino, L. Faoro, G. Falci, and R. Fazio, Phys. Rev. Lett. \textbf{88}, 228304 (2002).
%
\end{thebibliography}
\end{document}